\documentclass[12pt,a4paper]{article}
\pdfoutput=1
\usepackage{jheppub}
\usepackage{amsmath,amssymb,amsbsy,amsfonts,latexsym,graphicx,hyperref}
\usepackage{color,array,subfigure}

\newcommand{\be}{\begin{equation}}
\newcommand{\ee}{\end{equation}}
\newcommand{\bea}{\begin{eqnarray}}
\newcommand{\eea}{\end{eqnarray}}
\newcommand{\beq}{\begin{equation}}
\newcommand{\eeq}{\end{equation}}
\newcommand{\beqa}{\begin{eqnarray}}
\newcommand{\eeqa}{\end{eqnarray}}

\newcommand{\abs}[1]{\left| #1 \right|}

\renewcommand{\Re}{{ \rm Re}}
\renewcommand{\Im}{{ \rm Im}}

\title{Unitarity bound violation in holography and the Instability toward the Charge Density Wave }
\author[a]{Geunho Song,}
\author[b]{Yunseok Seo,} 
\author[a]{and Sang-Jin Sin}
\emailAdd{sgh8774@gmail.com}
\emailAdd{yseo@gist.ac.kr}
\emailAdd{sangjin.sin@gmail.com}
\affiliation[a]{ Department of Physics, Hanyang University, Seoul 133-791, Korea }
\affiliation[b]{ GIST College, Gwangju Institute of Science and Technology, Gwangju 61005, Korea }
\abstract{
We study the spectral function of holographic fermions with Pauli    term
and find  that when   bulk   mass goes  beyond the   unitarity bound  there is  an instability with  tachyonic dispersion relation. Based on the the linear density dependence of wave vector involved,  we suggest that the instability is toward the charge  density wave(CDW) whose wave vector  can be read off from the position of  the tip of k-gap. We  point out the similarity between the unitarity violation  in this model and the 'Nesting' as a mechanism of CDW. }

\subheader{\today
  }

\arxivnumber{}

\begin{document}

\maketitle

\section{Introduction}  

Instability of a theory is the indication that  the physical system is not   described by the correct degrees of freedom in the parameter regime it happens. It also plays a role of a messenger telling us   that  a new phase is  ready  there.  The simplest  example is the interacting scalar theory with its potential term $V[\phi]=r\phi^{2}$.  For  $r<0$, the fluctuation around $\phi=0$ is tachyonic one,   leading  to an instability.  
The system  develops a new vacuum and the prescription to the symptom  is to add $\phi^{4}$ term to  $V[\phi]$ by considering higher interaction.
 In a conformally invariant system, one of the stability requirement is the conformal unitarity bound.  Often the unitarity violation is associated with the presence of the `tachyon', signaling an instability, which is natural because unitarity violation means that the degrees of freedom is sinking or emerging rather than being conserved.  In fact,  whenever a phase transition happens,  
physics is described by a new degree of freedom  which is not given in the original hamiltonian. Also going from UV to IR fixed point is not 
a unitary process as the c-theorem of CFT says.  
At the level of the effective theory the old  degrees of freedom  were destroyed and new degrees of freedom were created around the phase transition point,  therefore in a sense  unitarity is doubly violated  there. Therefore it is valuable  to intentionally violate the unitarity bound  to see   what is going   beyond the regime and try to get information about the new phase.  
Often it gives us useful information on the nature of the instability and the  character of the new vacuum. 
 
On the other hand, understanding the mechanism of CDW in strongly interacting system attracted  much attention\cite{WHANGBO_1991,Ling:2014saa,Amoretti:2017frz}  due to its appearance  in the under-doped high Tc materials but most of the work were done  by transport calculation. Therefore it would be interesting  to consider it  in terms of  of the spectrum of fermions, which is the most basic information on the system analogous to the band structure of weakly interacting system, where the mechanism of CDW  instability is identified as the {\it nesting},  a phenomena that happens when a finite fraction of the Fermi sea (FS) is mapped to another part of FS by a single momentum vector {\bf  Q}.  There, the back scattering is  singular due to the divergently effective scattering sources available at each point of nested region. 
As a consequence,  the original  ground state becomes  unstable  and the system must move to a new ground state. 

 In this paper we study the spectral function of fermions in the presence of the  Pauli term ${\bar\psi}\Gamma^{\mu\nu}\psi F_{\mu\nu}$  in the holographic context in the regime where the bulk mass $m$  goes beyond the unitarity bound $|m|\leq1/2$, and find out that there is k-gap phenomena where the dispersion relation is that of tachyon. So here, violation of the unitarity bound  is associated with an instability.   We suggest that  such instability has strong similarity with the appearance of the charge density wave (CDW) through the nesting mechanism. 
The point is that the singularity in the scattering amplitude  in the presence of the nesting  leads to  the divergence of density of state which can be interpreted as the emergence  of new degrees of freedom   not encoded in the original Hamiltonian and 
this is very similar to going into the regime beyond the unitarity bound.  

The unitarity violation  can correspond to  many different physical situations depending on the interaction involved. 
The reason why the unitarity violation with Pauli term interaction is associated with CDW instability  are as  follows:   
\begin{enumerate}
\item Pauli term is based on the presence of the charge density described by  $A_t$.
\item Pauli term together with    bulk mass outside the unitarity bound   generates the gap, while the minimal interaction of  $A_t$ with the fermion does not  for bulk mass.  
\item More quantitatively, the location of the tip of the k-gap  depends on the charge density for low density and it matches with the experiment. 
\end{enumerate}

Previously, the CDW instability in the holographic setup  has   been discussed  \cite{Ling:2014saa,Andrade:2017ghg} by introducing  the space dependent chemical potential at the boundary of AdS as an input. The question was whether such input  survive when it evolved into the core region of AdS.   
Here what we search an indicator signalling the presence of the CDW without introducing the inhomogeneity explicitly.

  \section{ Spectral function : the setup and review}
We   consider the fermion action in the   dual spacetime with  the dipole interaction\cite{Edalati:2010ge,Edalati:2010ww} 
\begin{align}
S_{D} =\int d^4 x \sqrt{-g}\, i \,\bar{\psi}\left( \Gamma^M {\cal D}_M -m - i  p\, \Gamma^{MN} F_{MN}  \right)\psi   + S_{\text{bd}},  \label{Eq:S_Dirac}
\end{align}
where  the covariant derivative is
\begin{align}
{\cal D}_M  \;\!= \partial_{M} +\frac{1}{4} \omega_{abM}\, \Gamma^{ab} -i q A_M.
\end{align}
For fermions,  we can not fix the values of all the component at the boundary,  and it is necessary to 
introduce `Gibbons-Hawking term' 
$S_{\text{bd}}$. The equation of motion which is 
defined as
\beqa
S_{\text{bd}} =  \frac{1}{2} \int d^dx \sqrt{h} \,\bar\psi \psi  =\frac{1}{2} \int d^dx \sqrt{h} \,(  \bar\psi_- \psi_+ + \bar\psi_+ \psi_- ), \label{Eq:S_bd}
\eeqa
where $h= - g g^{rr}$, $\psi_\pm$ are the spin-up and down components of the bulk spinors.   
 The former defines the standard quantization and the latter does the alternative quantization.
 The background solution  we will use is Reisner-Nordstrom black hole in asymptotic $AdS_4$ spacetime,
\begin{align}
ds^2 &= -\frac{r^2f(r)}{L^2} dt^2 +\,\!\frac{L^2}{r^2 f(r)} dr^2 +\frac{r^2}{L^2}d\vec{x}^2  \cr
f(r) &= 1+ \frac{Q^2}{r^4}-\,\!\frac{M}{r^3},~~~~~A=\mu \,\! \left(1-\frac{r_0}{r}\right),  \label{AdS4}
\end{align} 
where    $L$ is {\it AdS} radius, $r_0$ is the radius of the black hole   and   
$
Q= r_0 \,\mu,~ M= r_0(r_0^2 +\mu^2). 
$
The temperature of the boundary theory is given by
$
T={f'(r_0)}/{4 \pi} 
$
and it  is related to $r_0$  by  
$r_0 = ( 2\pi T +\sqrt{(2 \pi T)^2 +3 \mu^2})/3$.
  
Following \cite{Liu:2009dm}, we now introduce  $\phi_{\pm}$ by 
\beqa
\psi_\pm = {(-gg^{rr})}^{-\frac{1}{4}} \,
 \phi_\pm, \quad \phi_\pm = 
\left(
\begin{array}{ccc}
  y_\pm    \\
  z_\pm    \\
\end{array}
\right), \label{Eq:psi_pm}
\eeqa 
after Fourier transformation. 
  Introducing the $\xi_\pm $ by 
$ \xi_+ \!=i   {y_-}/{z_+}, \hbox{ and }~  \xi_- \!= - i { z_-}/{y_+},$   
  the   equations of motion   can be given by 
 \beqa
\sqrt\frac{g_{xx}}{g_{rr}} \,\xi_\pm^\prime = -2m \sqrt{g_{xx}}\,\xi_\pm  + [  u(r)   -p \sqrt{g_{xx}} A_{t}'(r) \mp k]  + [ u(r) +p \sqrt{g_{xx}} A_{t}'(r) \pm k ]\xi_\pm^2. \label{eomxipm}
\eeqa
and the Green functions    for  $m<1/2$  can be written  as  
\beqa
G^R_\pm (\omega,k) = \lim_{r\to\infty} r^{2m}\xi_\pm(r,\omega,k). \label{greenless}
\eeqa 
Notice that   two components of the Green function are not independent: 
$ 
G^R_- (\omega,k)=G^R_+(\omega,-k).
$
Since  $G_{R}$ for  $m<0$ case, can be also obtained by $G_{R}\to -1/G_{R}$, 
   ${\tilde G_R}$, the Green function for the alternative quantization for $m>0$, is the same as  that for   $-m$ in the standard quantization: 
   \begin{align}
  {\tilde G_{\pm}^R(\omega,k;m)}= -1/G_{\pm}^R(\omega,k; m) =\,\!G_{\mp}^R (\omega,k;-m).
\end{align}  
In the presence of the dipole interaction, we get
\be
{ G_{\pm}^R(\omega,k;m,p)} = -1/G_{\pm}^R(\omega,-k; -m,-p).
\ee
The spectral function is defined as the  imaginary part of the Green function.  If we define  
$A_\pm(\omega,k)=\Im[G^R_\pm(\omega,k)]$    the spectral function is sum of them:
\be  
A=A_+(\omega,k)+A_-(\omega,k).
\ee 
It has been pointed out \cite{Gursoy:2011gz} that 
  the high frequency behavior of the spectral function  diverges like $\omega^{-2m}$ in alternative quantization which we take to maintain the positivity of $m$ in the interesting regime.     
 For the regime within the unitarity bound $|m|\leq 1/2$, we studied the fermion spectral function  \cite{Seo:2018hrc} using the method developed in \cite{sslee,Liu:2009dm,Faulkner:2009wj,Faulkner:2011tm,Faulkner:2013bna} in a holographic model
\cite{Zaanen:2015oix,Hartnoll:2016apf} to describe the Mott transition.  The  motivation of our previous study was to find a model interpolating the free fermion \cite{Cubrovic:2009ye,Cubrovic:2010bf} and Mott insulator  in the holographic setup\cite{Edalati:2010ge,Edalati:2010ww}. 
There, we restricted the bulk mass $m$ to $  \abs{m}\leq1/2$  for the conformal unitarity  and 
$m\geq 0$ for the normalizability of the spectral function
\footnote{
We stress that nothing change by taking standard quantization with negative $m$ because 
$A\sim \omega^{2m}$ and  $\Delta=\frac d2 + m $ while in alternative quantization these formula changes to 
$A\sim \omega^{-2m}$ and $\Delta=\frac d2 - m $. We work in  in alternative quantization just  for the ease of the presentation using positive $m$.}.

 \section{k-gap as an instability  toward the charge density wave}
 Below, we  will study fermion spectral function for $m>1/2$ where conformal unitarity is violated.  
In the non-interacting system, the dispersion relation is due to the spectral function given by the  a delta function. For interacting case, however,  the spectral function is a smoothed out, where the dispersion relation is indicated at most by the resonant peaks. We may call it a fuzzy dispersion relation. It turns out that  when the Pauli term is large enough, the  dispersion relation is given by   
  \be 
\omega^{2}- v^{2} k^{2}=M^{2}, \quad {\rm with}  \quad M^{2}<0,\ee
which is called as $k$-gap instead of the usual  gap  with  $ M^{2}> 0$. 
 That is, the spectrum is  tachyonic, indicating the  instability of the vacuum.  We will see that this actually happens in our model in the regime $m>1/2$.

\subsection{ Appearance of k-gap and and instability}
\begin{figure}[ht!]
\centering
        \subfigure[$m=0.75$]
   {\includegraphics[width=3.6cm]{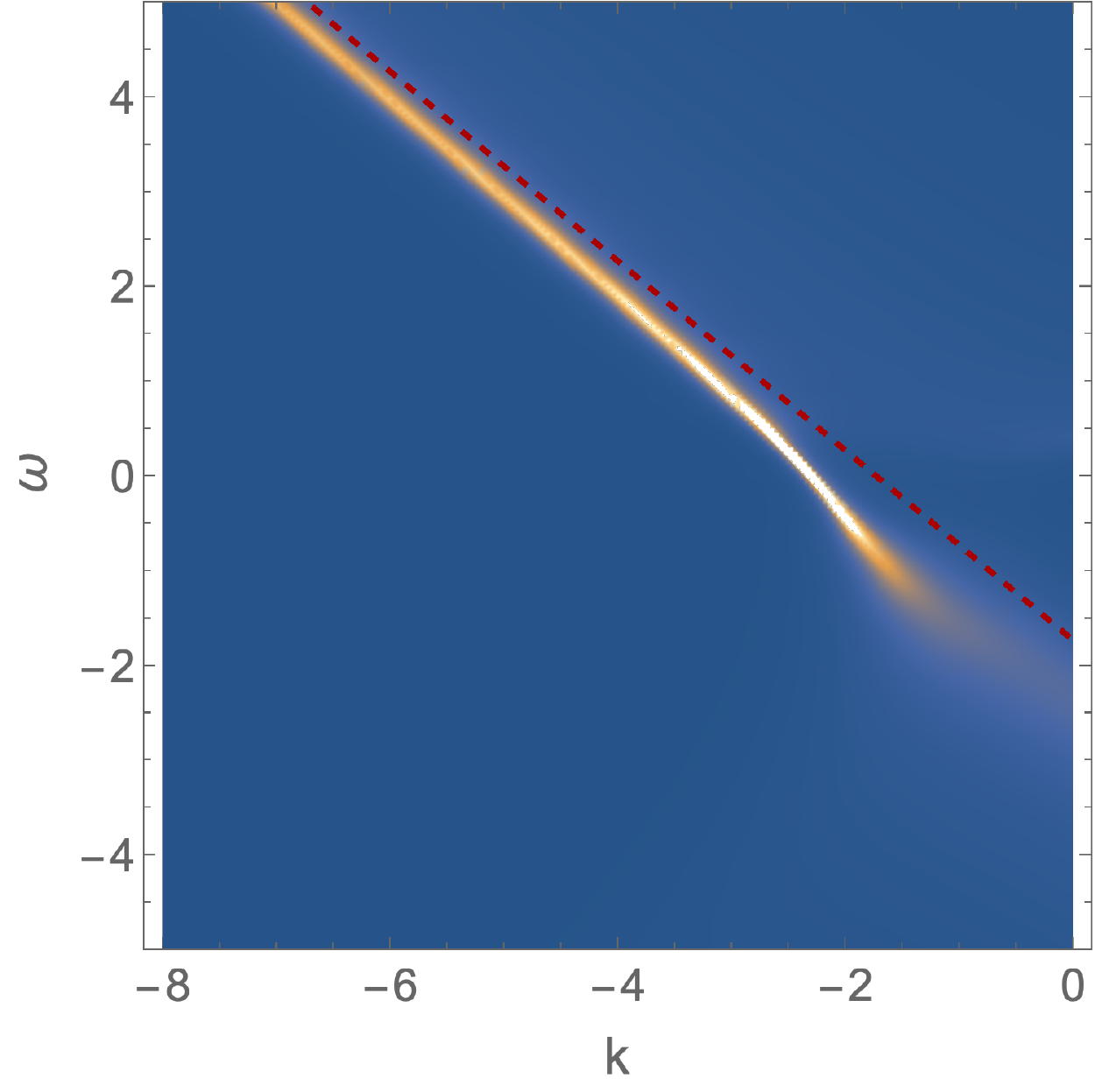}  }
       \subfigure[$m=0.95$ ]
   {\includegraphics[width=3.6cm]{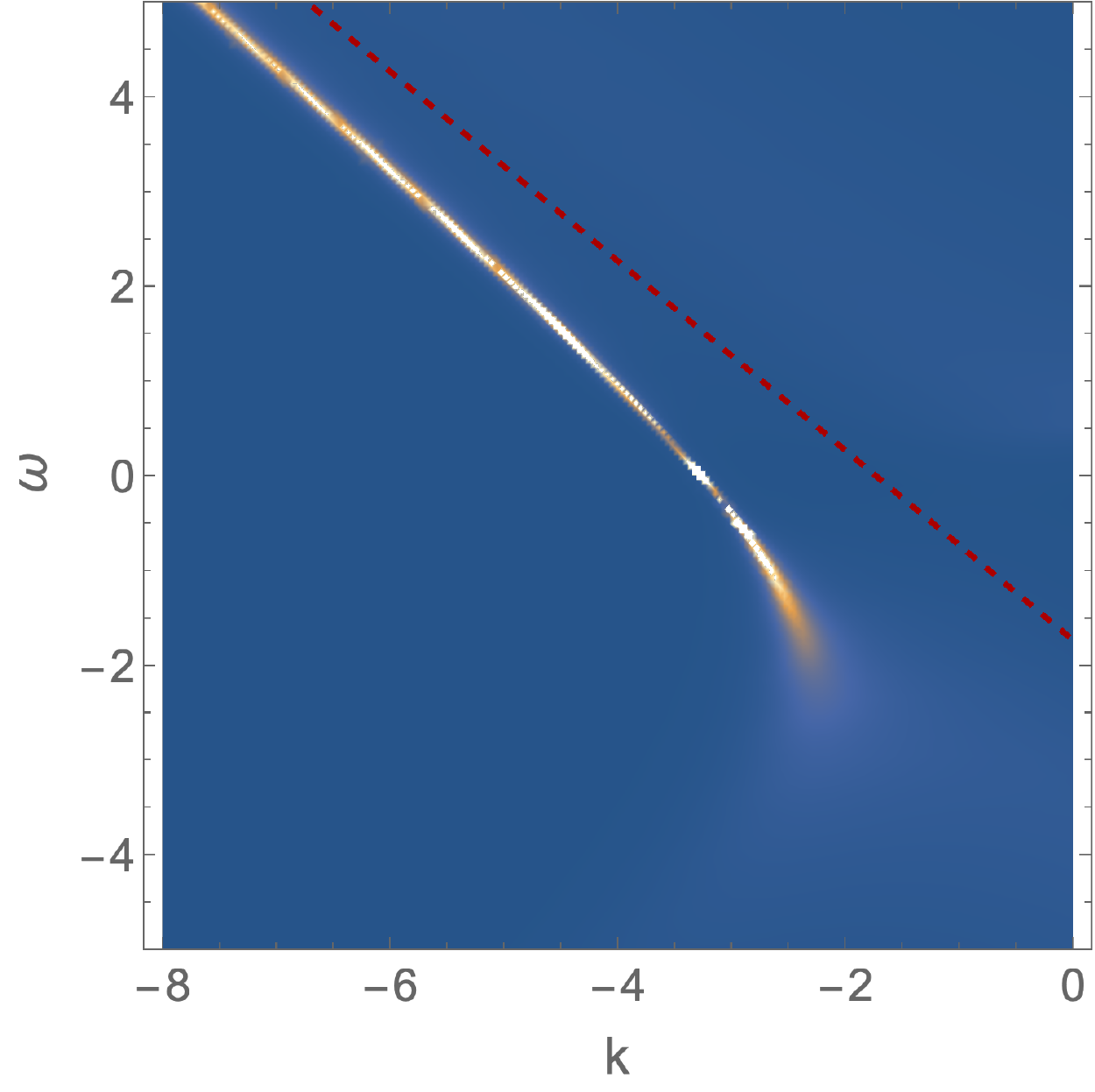}  }
     \subfigure[$m=0.999$ ]
   {\includegraphics[width=3.6cm]{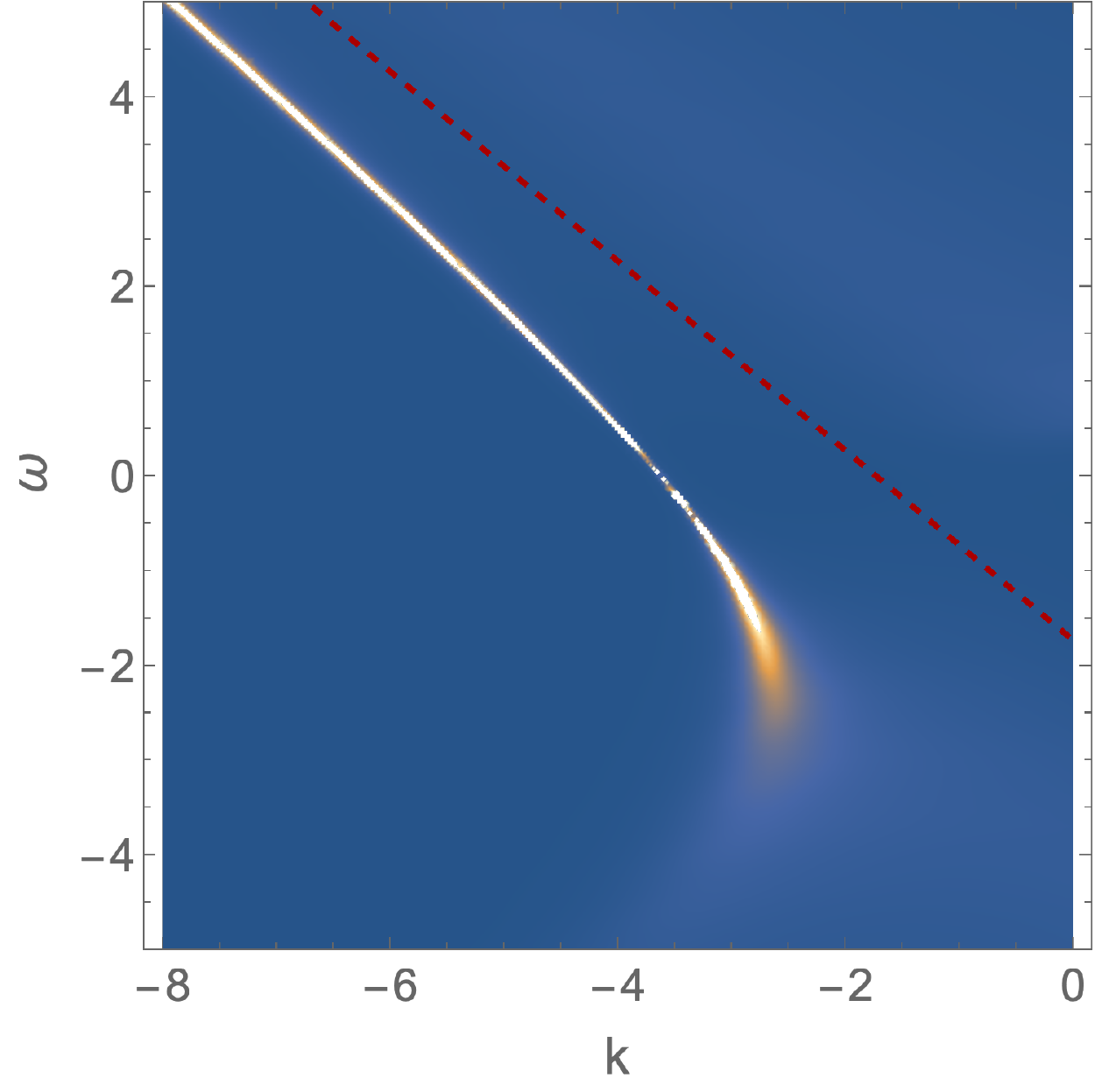}  }
      \subfigure[ $m=1.2$ ] 
    {\includegraphics[width=4cm]{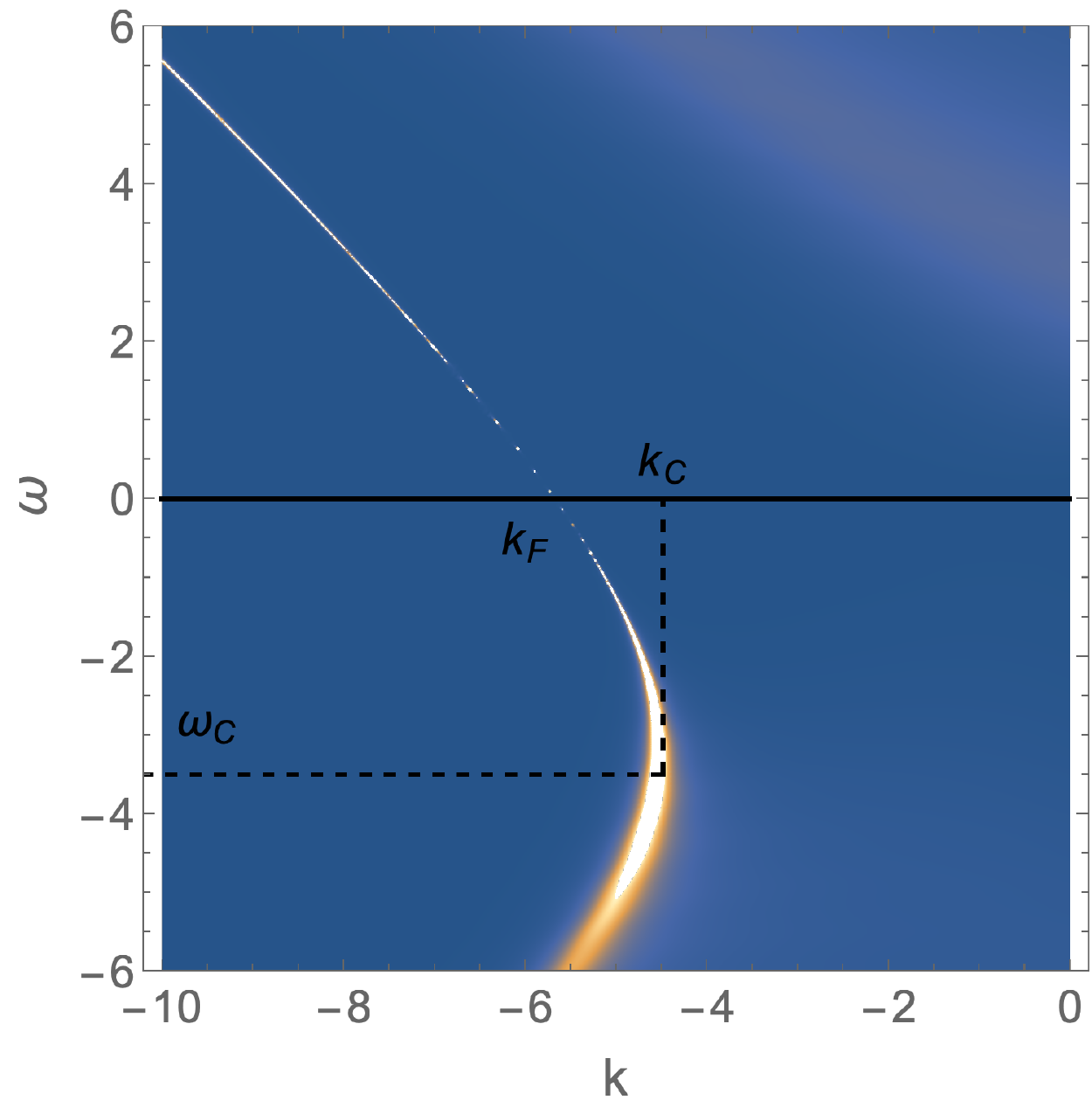}  }
    \caption{$m$-Evolution of the spectral density for   $p=1/2$. We have k-gap only if $m>m_c\simeq 0.96$. (d)  also shows the definition of  ($k_{C}, \omega_{C}$) as the tip position.
 }     \label{fig:evolution}
\end{figure}

\begin{figure}[ht!]
\centering
   \subfigure[  $m =0.3$]
  {\includegraphics[width=3.5cm]{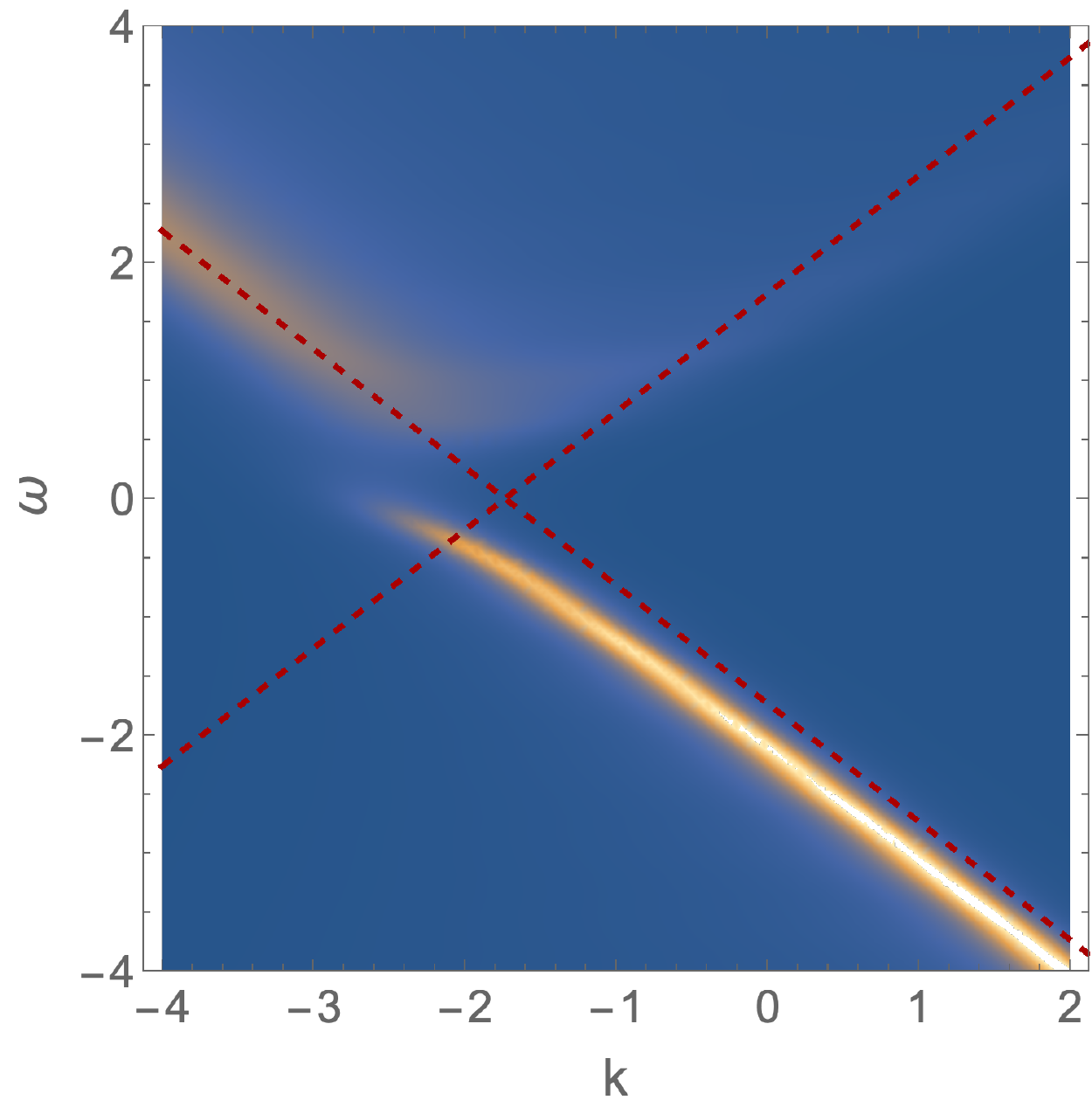}  }
       \subfigure[ $m =0.4$ ]
   {\includegraphics[width=3.5cm]{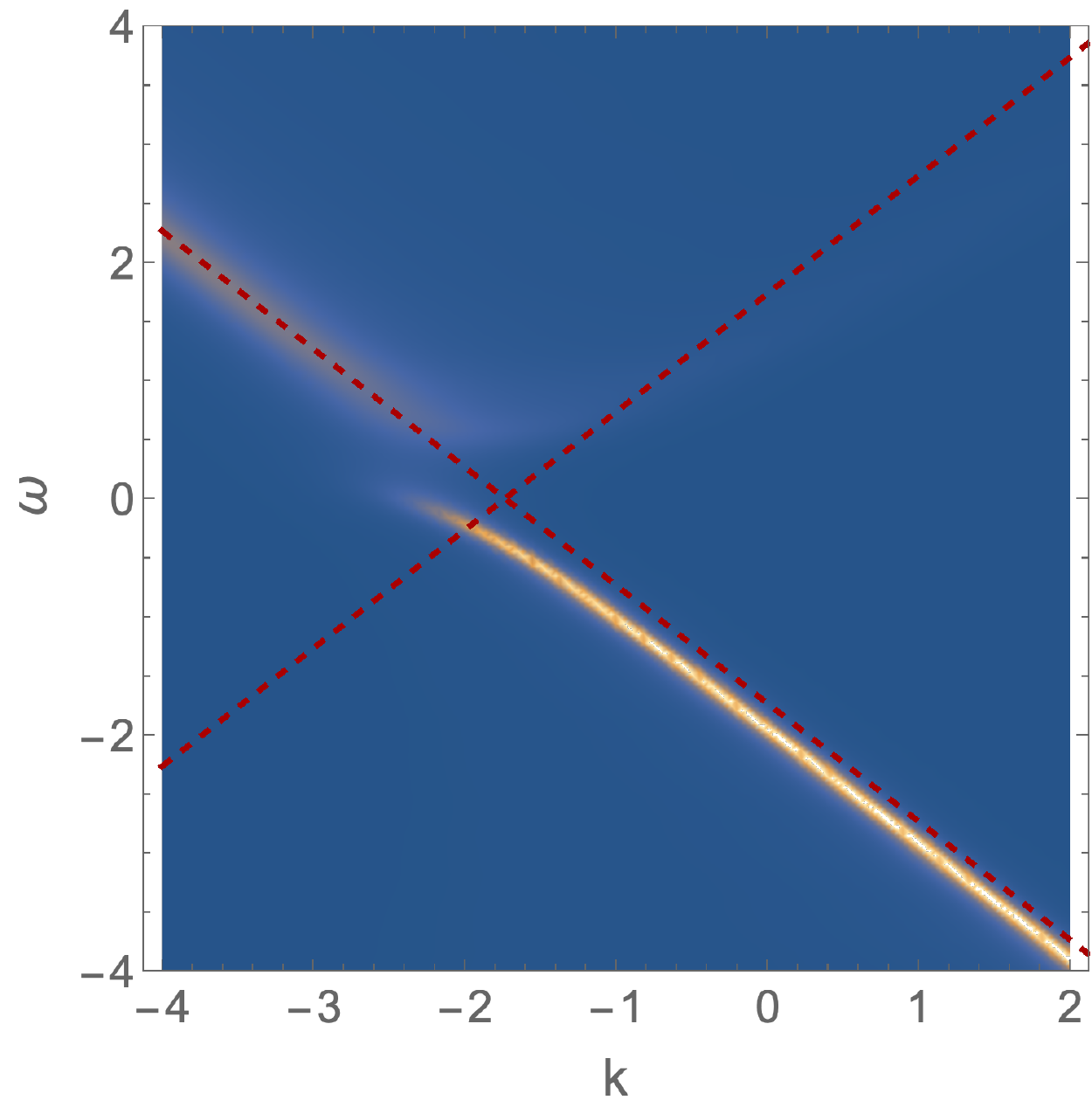}  }
        \subfigure[ $m =0.6$]
    {\includegraphics[width=3.5cm]{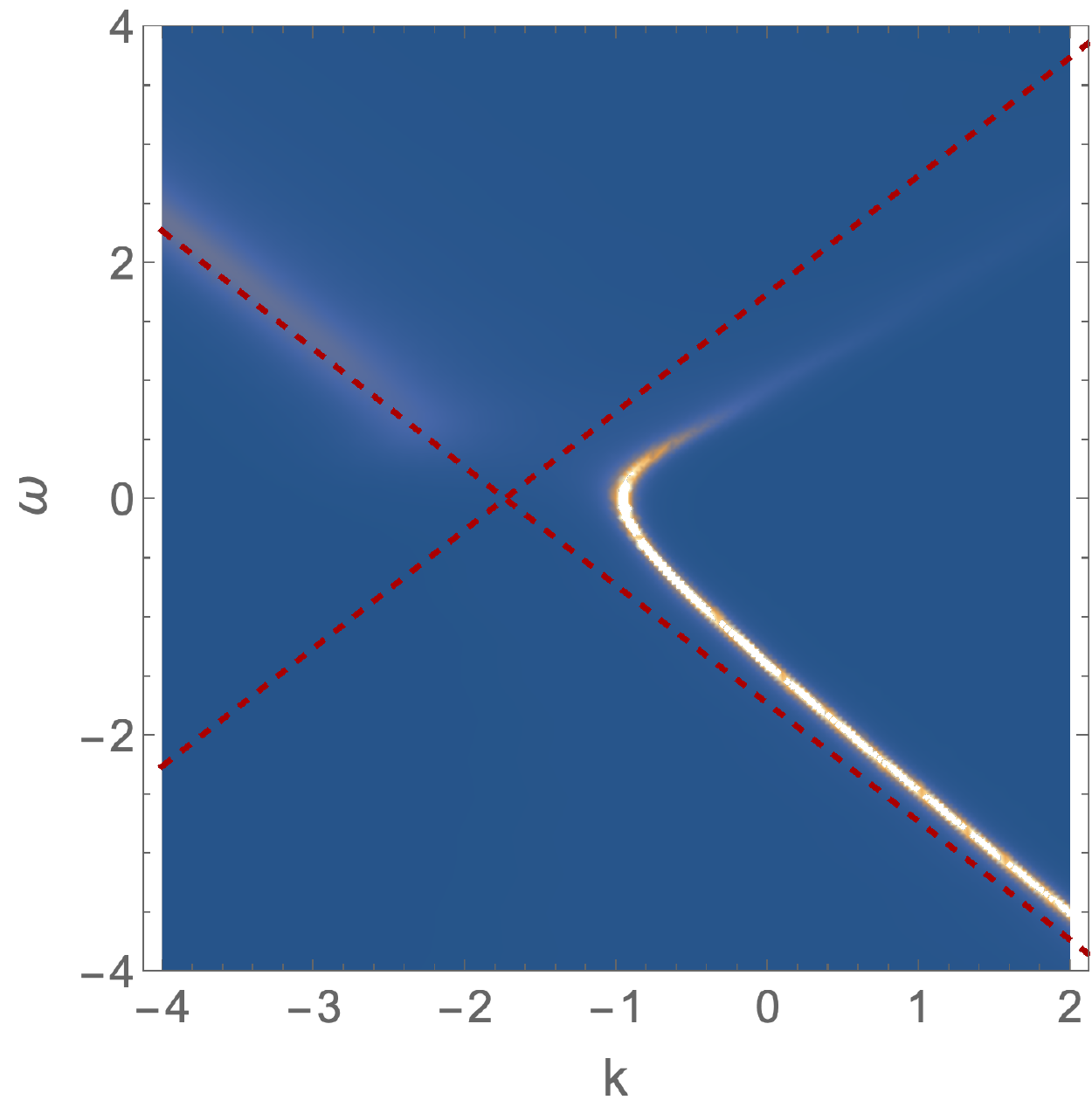}  } 
        \subfigure[$m=0.7$ ]
   {\includegraphics[width=3.5cm]{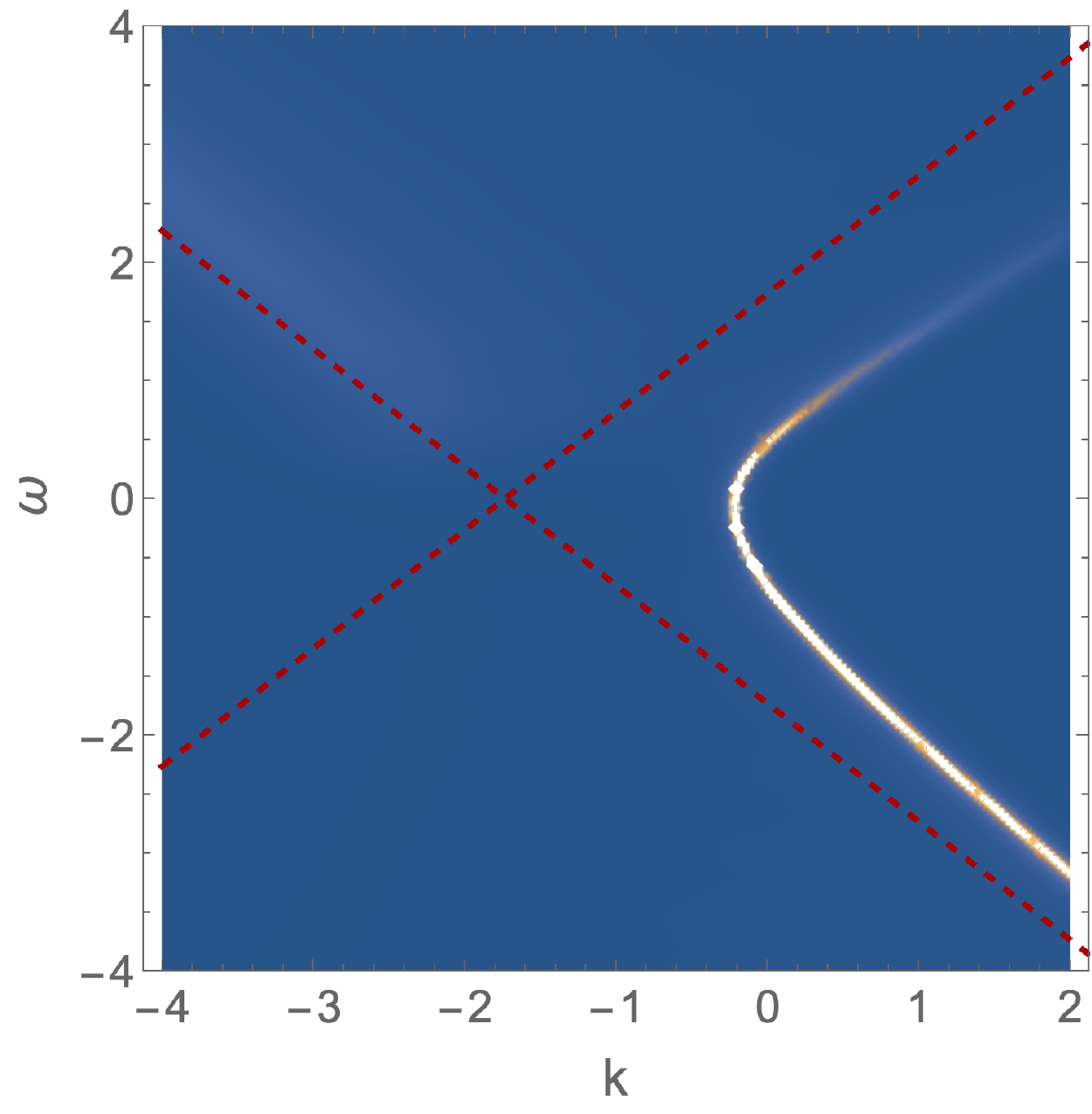}  }
\caption{$m$-Evolution   across the $m = 1/2$ with    $p=2$ fixed.      k-gap appears for any $m>1/2$.   
Red dotted line denotes IR light cone.  When $m=1/2$, spectral density  coincides with IR light cone }
      \label{fig:evol1}
\end{figure}

The $m$-evolution of the spectrum can be traced by the zero of $C$. 
First we consider  $p<1$. 
For unitarity bound $m<1/2$, 
it was shown that the system is in  gapless phase 
\cite{Seo:2018hrc}. 
Here we ask for  $m>1/2$ and the results are shown in Figure \ref{fig:evolution}(a-c).  Notice that there is a critical value  $m_c\simeq 0.96$ where  that k-gap begins to appear. If k-gap shows up, there is a window of $k$ for which there is no value of $\omega$.  
 Figure \ref{fig:evolution}(d)  to  shows the definition of   $(k_{C}, ~\omega_{C})$ as the position of the tip of the hyperbola. 
Now we move to the case $p>1$. Here it is most interesting to see the evolutions  across the $m =1/2$, which  is drawn in Figure \ref{fig:evol1}.  The key result is that k-gap appears whenever $m$ is bigger than 1/2. 
That is, for large enough Pauli term, unitarity violation always gives 
instability.  
 This  is  the kind of the  phenomena expected in the regime where the unitarity bound is violated and indeed it happens. 

\begin{figure}[ht!]
\centering
    {\includegraphics[width=8cm]{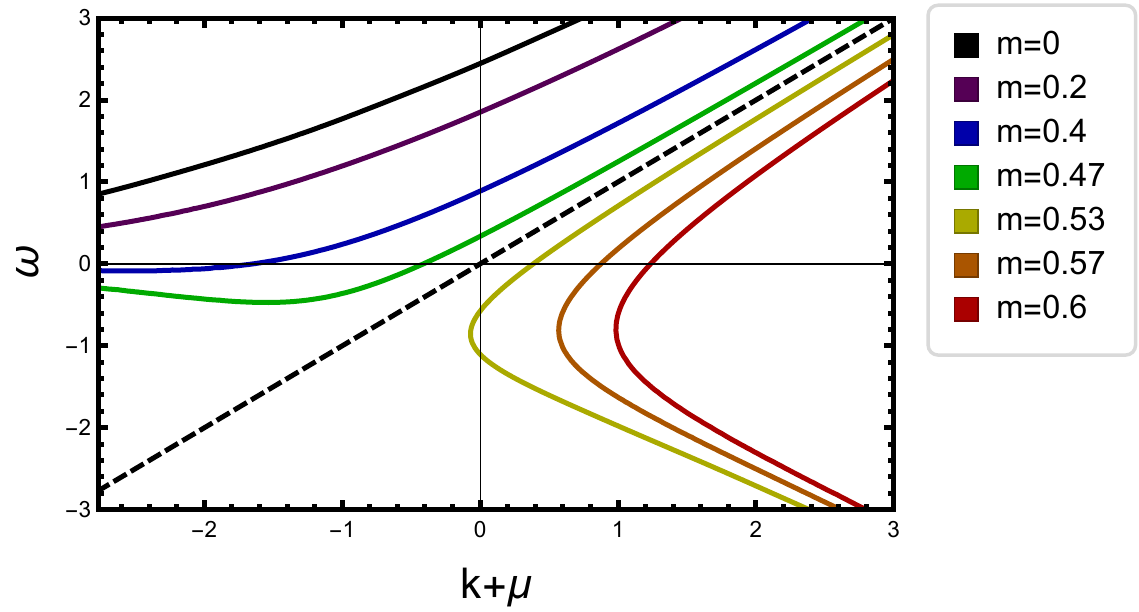}  }
    \caption{Mass evolution of spectral density at $p=5$. Dashed line denotes free fermion dispersion, $\omega = k+\mu$. }
      \label{fig:kgap}
    \end{figure}

   \vskip .1cm
Figure \ref{fig:kgap} collectively describes the evolution of the dispersion curve as we change the $m$  for $p=5$. 
Starting from $m=0$ which  belongs to the gapped phase, it arrives at the  free fermion like phase  at $m=1/2$, and finally it runs into the k-gapped state when $m>1/2$.  This is very analogous to   of the phenomenon that  appears in the recent work\cite{Baggioli:2018vfc,Baggioli:2018nnp} describing a transition in transverse phonon dispersion from solid to liquids.     
In drawing the Figure  \ref{fig:kgap},  we used the method locus of $\Re G^{-1}=0$,  which will be described in appendix. 


\subsection{Fate of the vacuum and the Charge Density Wave} 
\begin{figure}[ht!]
\centering
    \subfigure[Density dependece of $k_c$]
    {\includegraphics[width=4.5cm]{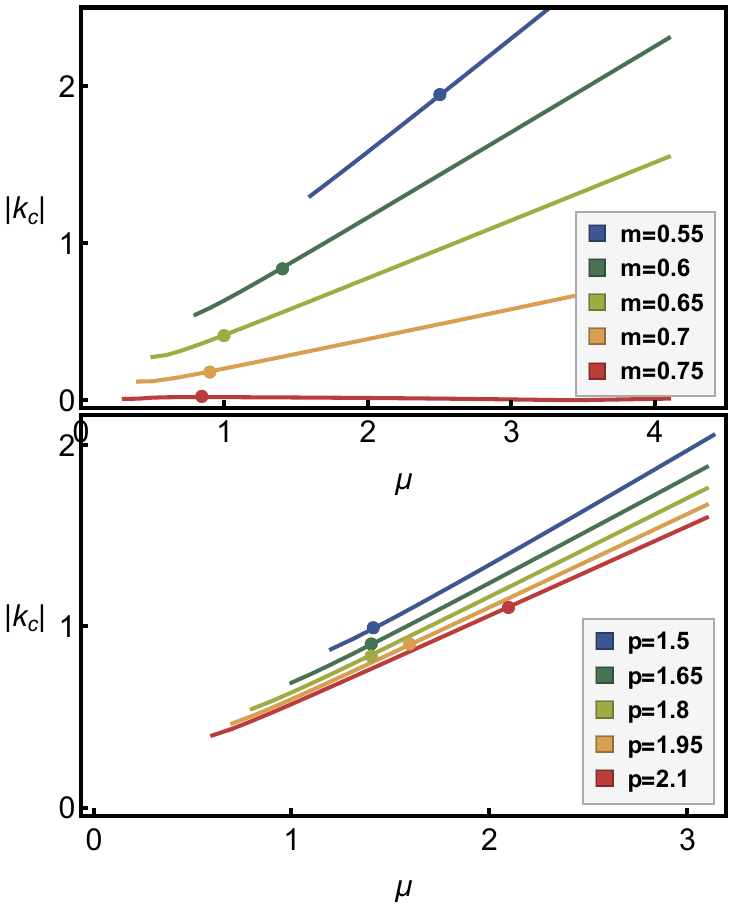}  }
        \subfigure[CDW vector vs doping]
    {\includegraphics[width=5cm]{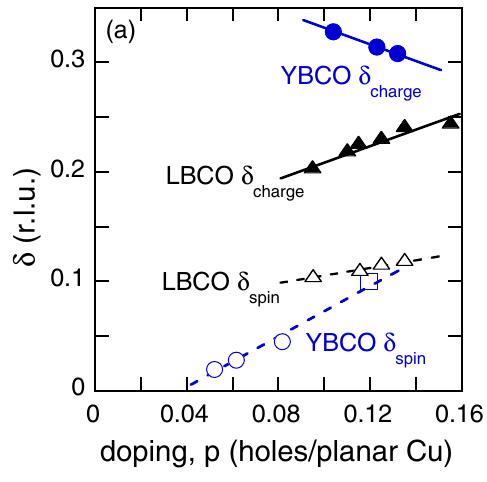}  }
   \subfigure[bending data]
  { \includegraphics[width=5.5cm]{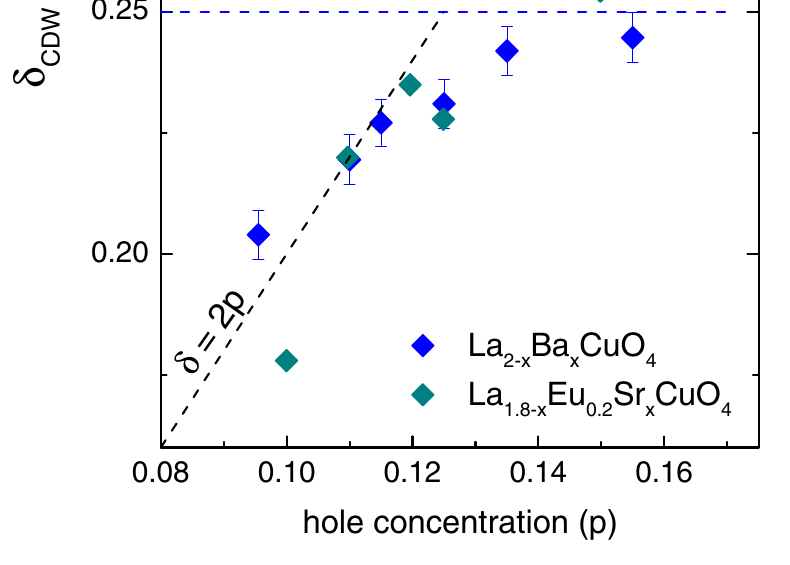} }
    \caption{ (a) Linear dependence  of $k_C$ in  $\mu$, suggesting that the instability is toward the charge density wave (CDW).  We choose $p=1.8$ (upper panel) and $m=0.6$ (lower panel). The dots are the location of $k_{C}^{*}$. 
    (b,c) The doping  dependence of  CDW vector $\bf\delta$. The figure (b) is reproduced  from     \cite{blackburn2013x} and figure (c)  is from     \cite{blanco2014resonant}.
    }
      \label{fig:linearity}
    \end{figure}

To  see the nature of the instability, we calculated  the density dependence of the position of $k_C$ and found that  there is a linear relation between them. See Figure \ref{fig:linearity}(a). 
The strong correlation between the charge density and the wave vector at the instability 
 indicates that the instability is associated to the charge density wave. 
The Figure \ref{fig:linearity}(b)(c) is reproduced from \cite{blackburn2013x}, where linear dependence of the charge density wave vector $\bf \delta$ in the doping which is comparable to our charge density $Q$, which is related to $\mu$ by  $Q =\mu r_0$  and $r_0 = ( 2\pi T +\sqrt{(2 \pi T)^2 +3 \mu^2})/3$.  Therefore for  $\mu/T<<1$,  the density and the chemical potential is  also linear each other.  Since the doping parameter should be interpreted as the charge density the linear dependence of wave vector in the doping parameter 
is consistent with our theory  at least for the low doping.

For higher doping or charge density, $r_{0}\sim \mu$ so that 
$Q\sim \mu^{2}$ as expected from the dimensional analysis or the non-interacting Fermi gas theory. Therefore our theory shows $k_{C} \sim \sqrt{Q}$.
Figure \ref{fig:linearity}(c) shows the a bit more recent data 
which shows bending of the data, which  is consistent with our theory although the authors of the paper \cite{blanco2014resonant} tried to fit with linear plot without success. 
Although we do not aim to detailed match of our theory with the data, it is encouraging to see that the general trend of data vs theory is consistent.  This is our main result. 

Notice that the linearity of doping and the wave vector is not a character of weakly interacting system but that of a strongly interacting one. Also if we consider the charge in our theory  as the 'conserved' spin, we would get the spin density wave instead of the CDW.   

What would be the physical mechanism to create such unitarity bound violation? 
In a theory with Fermi surface(FS), CDW is associated with the nesting\cite{Rice:1975aa,WHANGBO_1991}, a phenomena associated  with a shape of FS where a single vector $k_{Q}$ connects large parallel regions of a FS. 
In perturbation theory, it can lead   to divergent scattering amplitude through $ \int dk/(E_{k}-E_{k+q}) \cdots  $ for $q=k_{Q}$ and $k\in $ $\{$Nesting region$\}$.
In the situation where Pauli principle is relaxed due to strong interaction so that the degrees of freedom relatively deep inside the Fermi sea can be excited,  the nesting mechanism 
 can work even in the absence of the Fermi surface. 
 In more physical terms, 
 such divergence can be rephrased as  the appearance of large degrees of freedom  that can interact   effectively  at a special momentum transfer $q=k_Q$, which is very similar to the unitarity  violation in our theory. 
   Therefore  it is   natural to  identify the tip of the k-gap as the  density wave vector:
   $k_C=k_Q$.
 
\subsection{A measure  of instability}
\begin{figure}[ht!]
\centering
\subfigure[ Spontaneous CDW instability]
{\includegraphics[width=7cm]{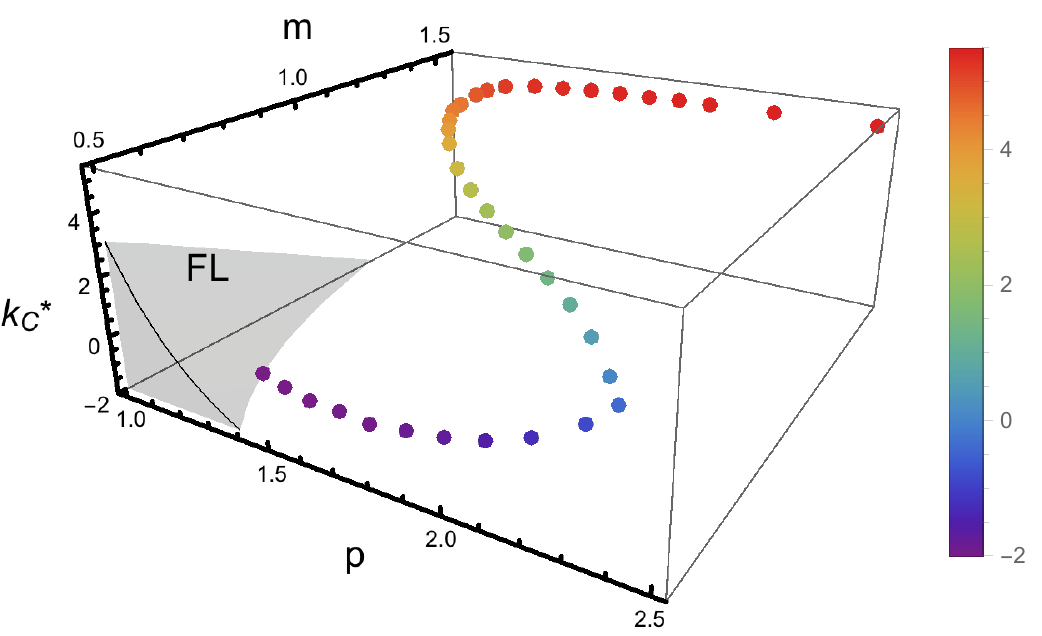}  }
\hskip1cm
 \subfigure[$m$-trajectory of  the tip position]
 	 {\includegraphics[width=6cm]{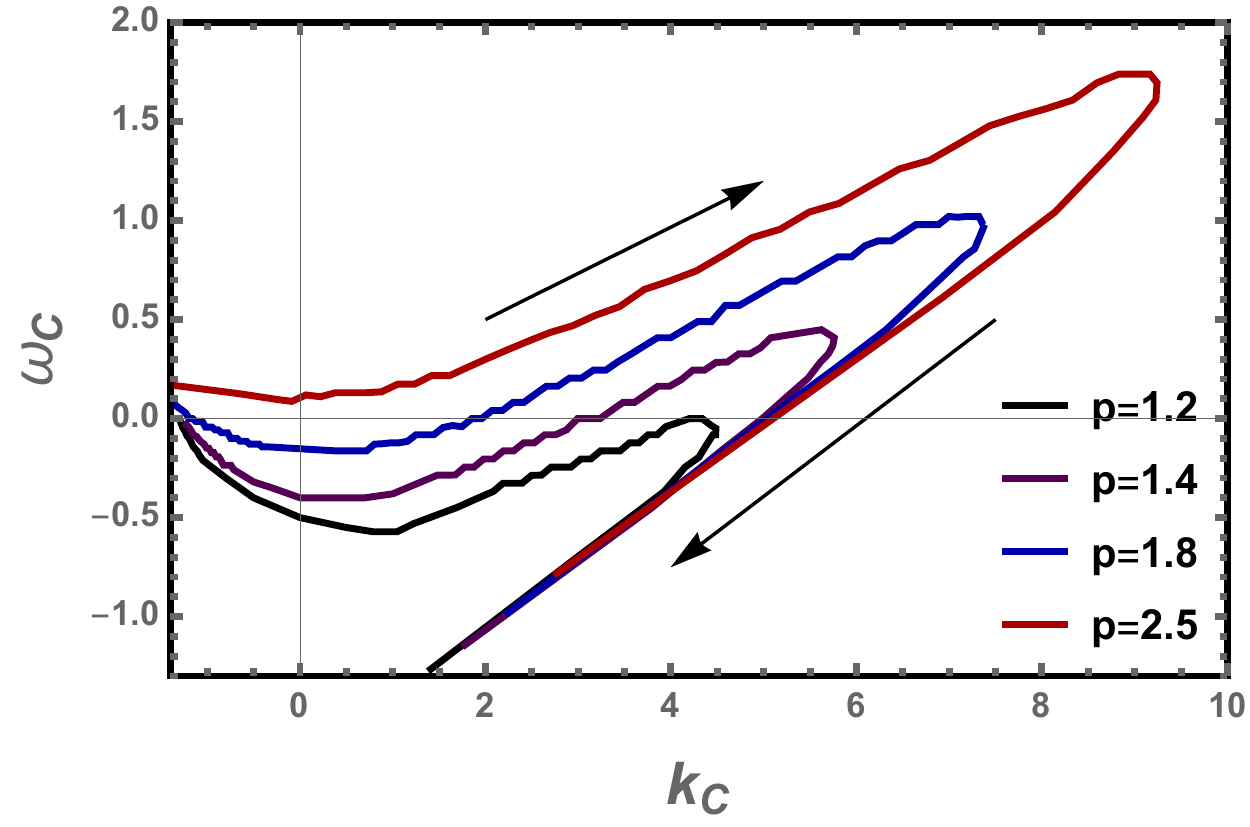}  }
 \caption{(a)  Trajectory of Spontaneous CDW: $D_{ins}=\infty$  along the dotted line.      (b) Trajectory of the tip  ($k_{C}, \omega_{C}$) drawn as a parametric plot as $m$ move: Arrow indicates direction of increasing mass.   }
      \label{fig:phase2}
\end{figure}

When the tip is located at the Fermi level, 
the instability is most vivid because   the  Fermi velocity diverges there.  The degree of super-luminosity, $v_{F}-1$,  can be used as a measure   of the instability: 
\be
D_{ins}= {v_{F}-1}.
\ee
When $D_{ins}=\infty$, the instability can happen spontaneously, while the system is barely unstable if $D_{ins}=0$. 
  Figure \ref{fig:phase2}(a) plot the trajectory of     $D_{ins}=\infty$   
 where   spontaneous $k$-gap generation appear in parameter space of   $(p,m,k_{C}^{*})$.   Here  $k_{C}^{*}$ is the value of  $k$ at the tip of the k-gap and we interpret it as the wave vector of charge density wave. 
 For any desired criterion $D_{0}$,  $D_{ins}>D_{0}$ define a tubular neighborhood along the trajectory where charge density wave is   formed with easiness $D_{0}$.  
 
Figure \ref{fig:phase2}(b) shows  $m$-trajectory    of the tip $(k_C, \omega_C)$ in  the momentum space for a few different values of $p$.  In the figure, the arrow indicates the direction of  increasing $m$. 
There are three different classes of trajectories according to 
the range of $p$.
\begin{enumerate}
\item  For $p < 1.3$,  $\omega_C$ is always negative and hence there is no spontaneous $k$-gap generation.  
\item  For $1.3 \le p \le 2.3$,  $\omega_C >0$  if  $m\simeq 1/2$. The   tip moves down as  $m$ increases and passes Fermi level where the Fermi velocity diverges. If we increase mass further, then the tip position moves up and down  so that the tip position pass $\omega=0$ line three times in total.
\item For $p > 2.3$, the position of tip starts from $\omega > 0$ region and pass $\omega=0$ line for $m \sim 1.5$.
\end{enumerate}
 
Notice  that  in the regime $p<1, m<m_c$ there is a region where  
k-gap or ghost spectrum does not appear in spite of the unitarity violation.  
See the shaded region of  \ref{fig:phase2}(b).  
This may be the fermionic analogue of the phenomena found   in ref. \cite{Andrade:2011dg} where it is shown for the scalar that there is a  ghost free region in spite of the violation of the stability bound.

\section{Discussion} 
In this paper, we calculated the spectral density of probe fermion 
with Pauli term above  the unitarity limit and found a characteristic k-gap for most parameter regime which   indicate that  the system has  instability. 
To show that the  toward the density wave, we calculated the density dependence of the tip of the k-gap. It turns out that it agree with the doping  dependence the wave number for physical system with CDW instability. 
This suggest that we can look for the spectrum and read off the wave number of the CDW by looking at the position of the tip of the k-gap.  

To discuss the physical mechanism for the unitarity violation causing the k-gap and CDW,  
  we point out that there is a similarity of unitarity bound violation  and the nesting phenomena:   in both cases   one can observe the appearance of large degrees of freedom that can interact very effectively at a special momentum deliverance.  
  We   emphasize that our method   did not  introduce  explicit inhomogeneity. 

We  speculate  that out of all possible fermion coupling $B_{{\mu\nu\cdots}}{\bar\psi}\Gamma^{{\mu\nu\cdots}} \psi$ with $B_{{\mu\nu\cdots}}$ being a linear function of $A_{t}$ and its derivatives, Pauli term is the only one that can generate the gap.

   \vskip .5cm
 {\bf Note added} After this work is uploaded to archive, we were informed that similar related 
 idea were discussed in recent papers \cite{Amoretti_2018,Amoretti_2018_2, Musso:2018wbv}, where low energy scaling dimension was gravitationally calculated as a function of momentum and instability was identified as breaking its reality.

 \vskip 1.5cm
\appendix
{\bf \Large Appendix}
 \section{Tracer of spectral function peak}
\begin{figure}[ht!]
\centering
    \subfigure[$m=0.5$, $p=0$ ]
    {\includegraphics[width=45mm]{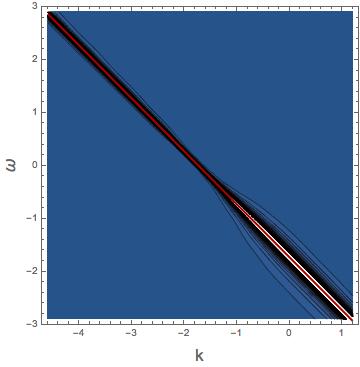}  }
        \subfigure[$m=0.5$, $p=5$ ]
   {\includegraphics[width=45mm]{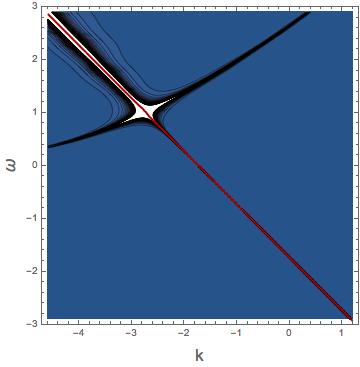}  }
       \subfigure[$m=0.47$, $p=5$ ]
   {\includegraphics[width=45mm]{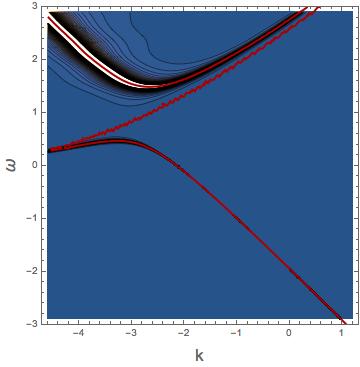}  }
    \caption{The zero of $C$ (red line) vs. the peak of spectral function. We use unsymmetrized $G_2$.
    (a) Here zero of $C$ overlap with the pole of the Green function. 
    (b) The new branch of dispersion curve created by the interaction and it is not followed by the 
    zero of $C$. 
    (c) Decreasing of $m$ from 1/2 deforms  the original dispersion curve is
and  the zero of $C$  follows  it,  of    but not the new branch. }
      \label{fig:specTOre}
\end{figure}

 We used locus of $C=0$ method in drawing Figure 3. 
Here we want to  describe it. If we write the retarded Green's function  as $G_{R}^{-1} = {C - i D}$ with real $C$ and $D$, 
\begin{align}
A = Im G_{R} = {D}/{(C^2 +D^2)}.
\end{align}
For finite $D$, the spectral density has its maximum at $C=0$. Therefore, one can easily check peak of spectral density by looking zero of $C$. It depends on the value of $D$ also.  When $C=0$, the spectral density becomes $A = 1/D$, hence it  is suppressed when  $D$ is large. 
To test how useful this idea is,  we calculate the zero of $C$ for $m<1/2$ and compare with the spectral function. 
The comparison to the spectral density and the zero of $C$ are drawn in Figure \ref{fig:specTOre}, from which we can see that the
two  are well matched, although they do not exactly overlap. 
Features    of Figure \ref{fig:specTOre}  are described below.
\begin{enumerate}
\item 
Figure  (a): At $p=0$, the peak of spectral function is same as zero of $C$.  
\item 
Figure   (b): The interaction generate new band along  the maximum of $D$, 
Out of $C=0$ line, $C>>D$ so that $A \sim D/C^2$.  
\item 
Figure   (c): For $m$ off $1/2$, the  original linear dispersion curve deform and the zero of $C$ follow it.  peaks of both band are overlapped to the zero of $C$. But there is another branch of zero of C, which is the middle band in the figure, which is not realized as the peak of the spectral function.  
Along this band, $D$ has maximum value and  the peak is suppressed since $A \sim 1/D$. 
\end{enumerate}

\begin{figure}[ht!]
\centering
   \subfigure[  $m =0.3$]
  {\includegraphics[width=3.5cm]{mevolm3.pdf}  }
       \subfigure[ $m =0.4$ ]
   {\includegraphics[width=3.5cm]{mevolm4.pdf}  }
        \subfigure[ $m =0.6$]
    {\includegraphics[width=3.5cm]{mevolm6.pdf}  } 
        \subfigure[$m=0.7$ ]
   {\includegraphics[width=3.5cm]{mevolm7.pdf}  }
   \subfigure[  $m =0.3$]
  {\includegraphics[width=3.5cm]{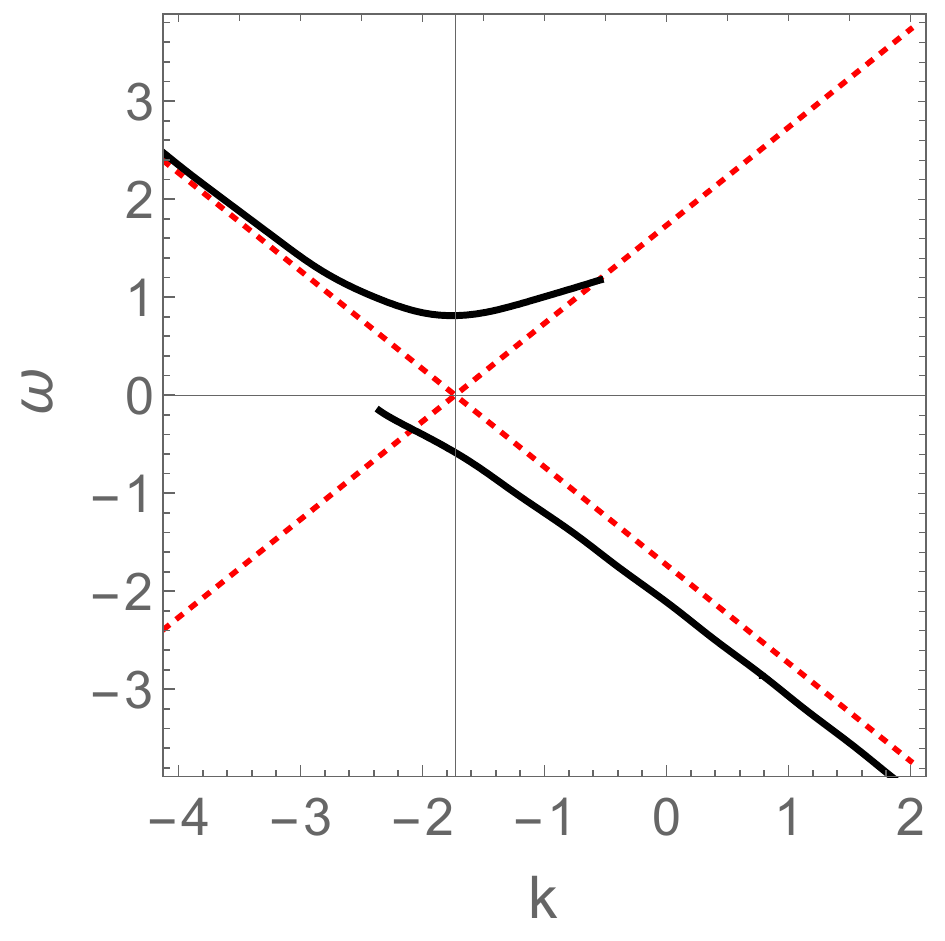}  }
       \subfigure[ $m =0.4$ ]
   {\includegraphics[width=3.5cm]{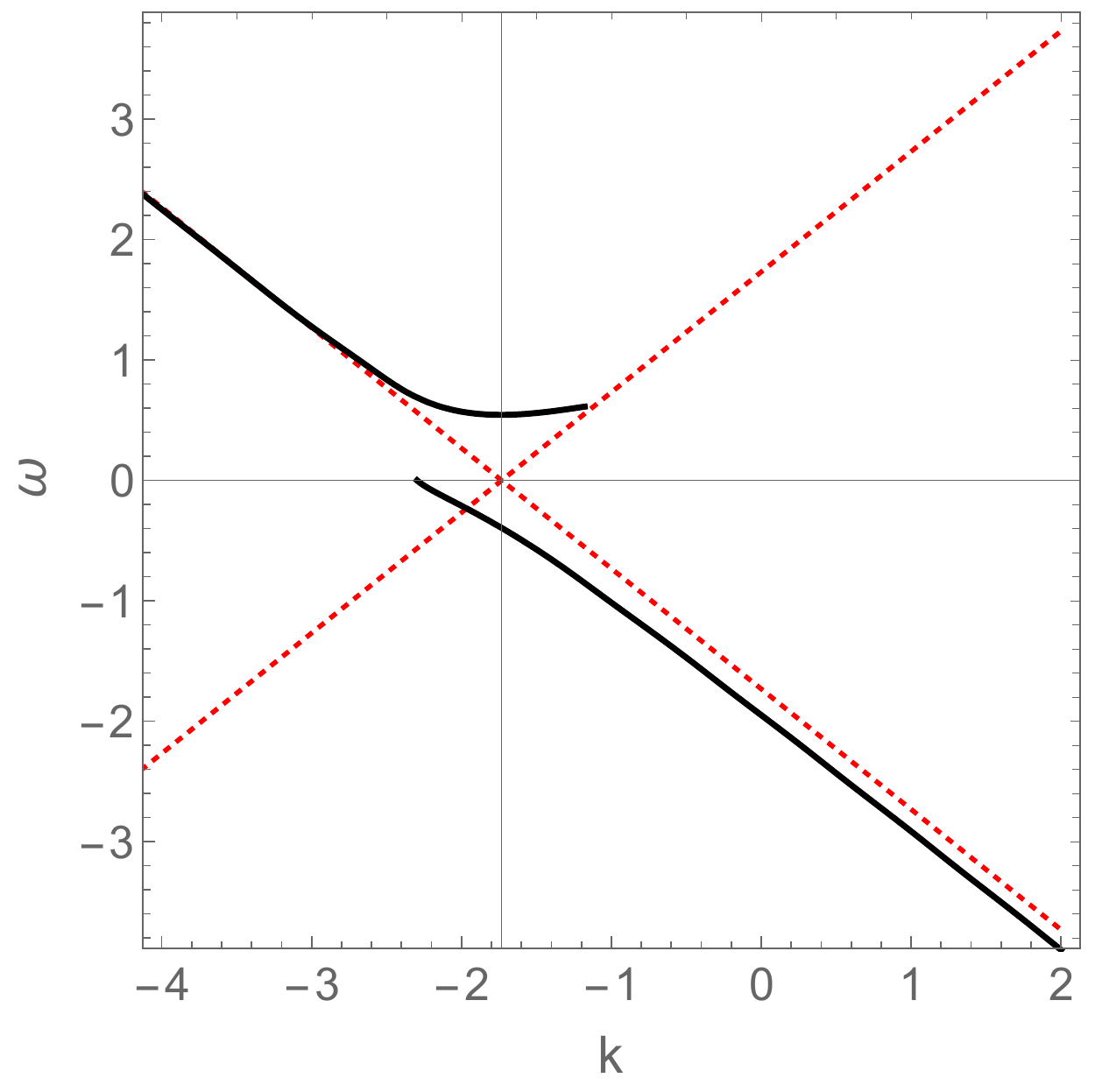}  }
        \subfigure[ $m =0.6$]
    {\includegraphics[width=3.5cm]{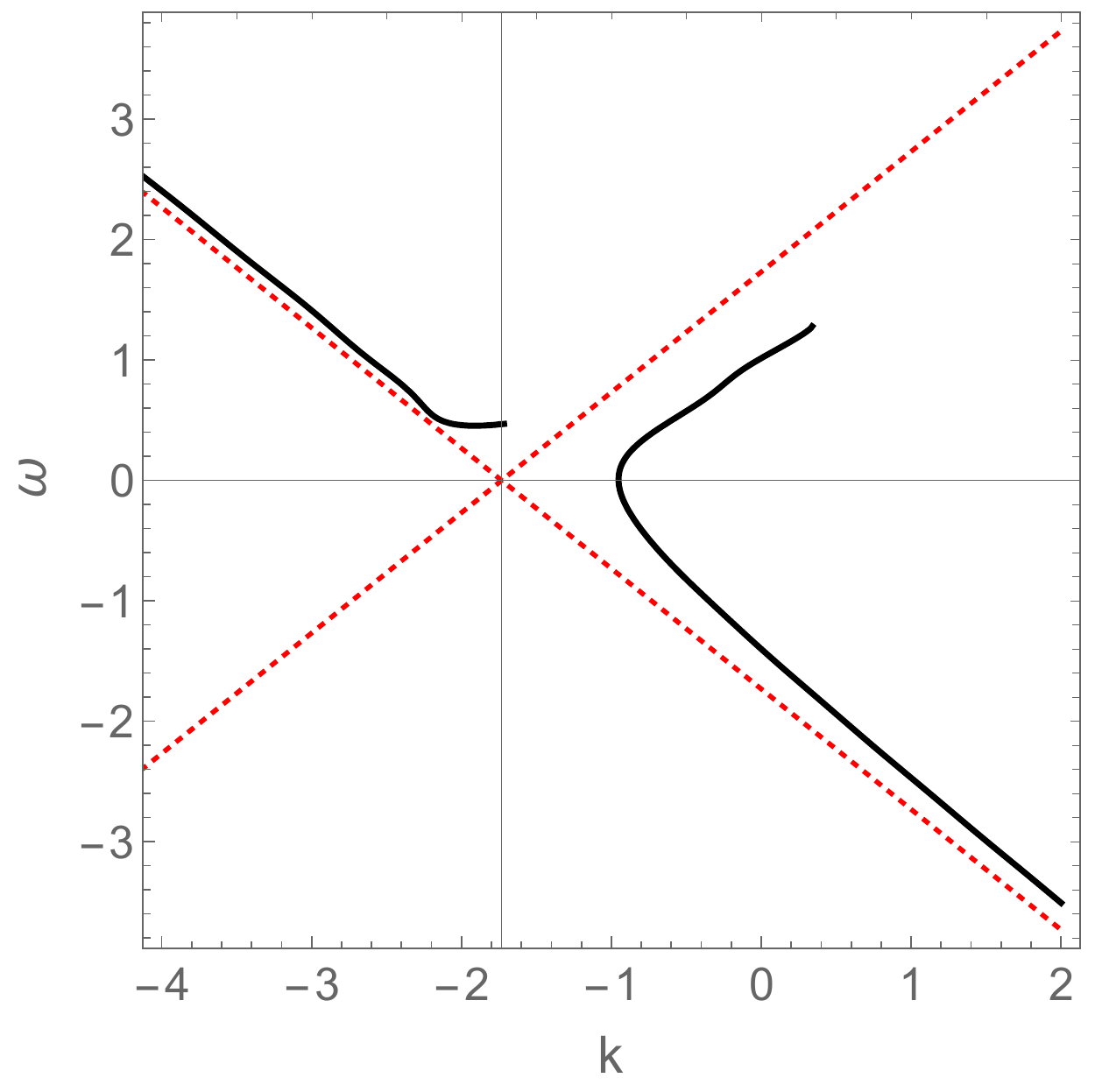}  } 
        \subfigure[$m=0.7$ ]
   {\includegraphics[width=3.5cm]{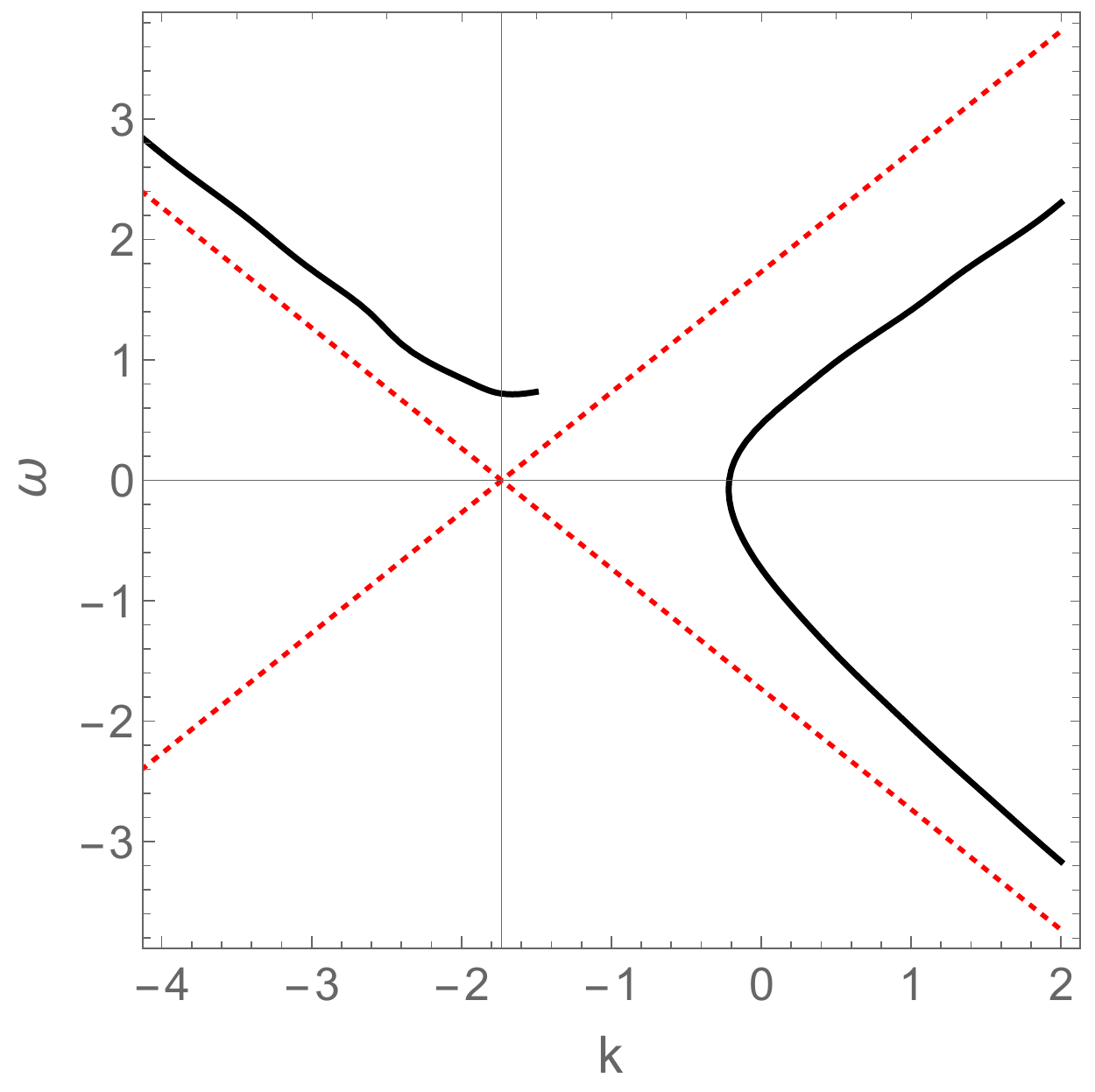}  }

\caption{Camparing the spectral density and locus of $C=0$ for  
$m$-Evolution   across the $m = 1/2$ with    $p=2$ fixed.   
Precise agreement is obtained.  }
      \label{fig:evol11}
\end{figure}

In Figure  \ref{fig:evol11}, we give some other comparison to test the method. We compared the spectral density and the locus of $C=0$ for  $m$-Evolution   with    $p=2$ fixed.   As one can see, almost precise agreement is obtained between the two. 
Summarizing,   the real part of $ G^{-1}$,   contains  the essential  information for   the peak of spectral density.  

\section{Phase diagram}
 Once we know that  the new vacuum has charge density wave,  
 can we figure out more details about the phases  of the system after CDW transition happen? 
      \begin{figure}[ht!]
\centering
   \subfigure[Phase diagram]
    { \includegraphics[width=7cm]{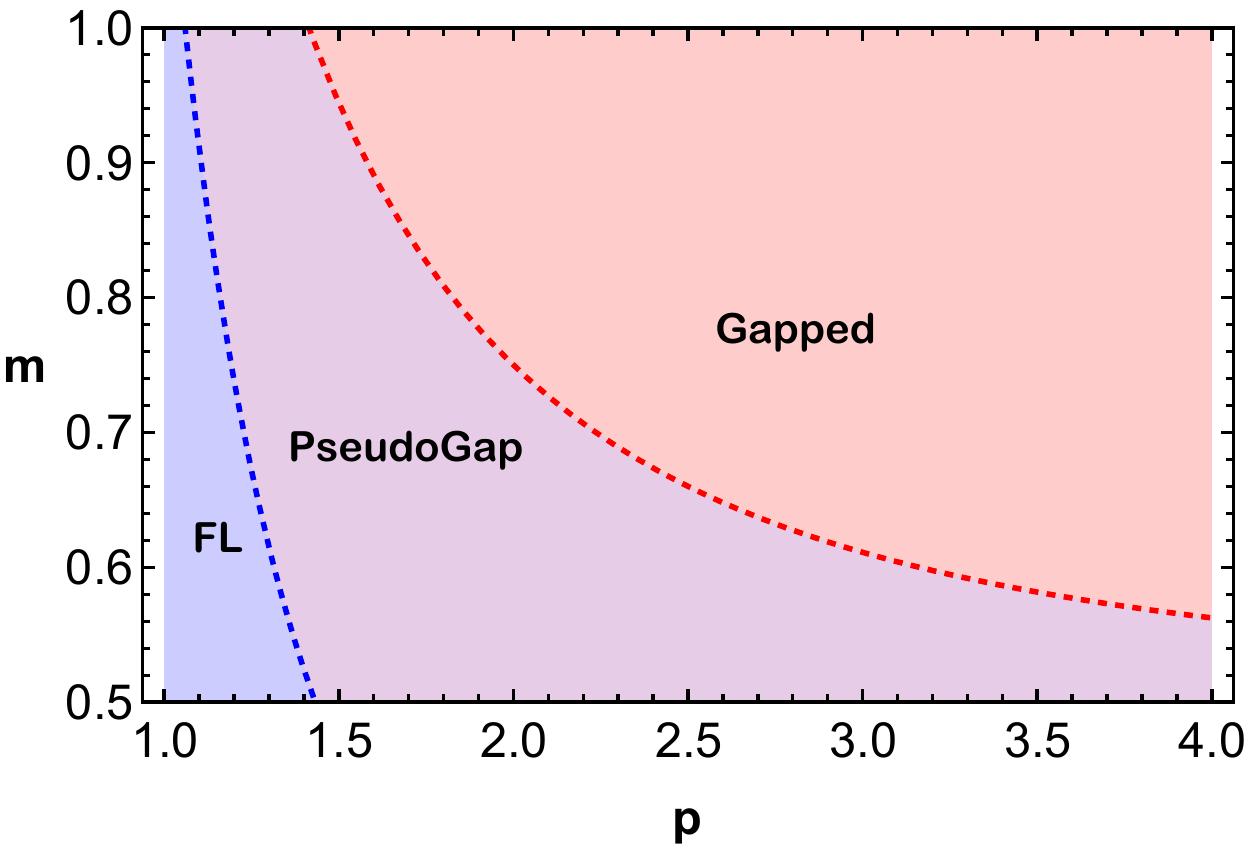} }
    \caption{ Phase diagram for $0.5<m<1.0$ with      the unstable k-gapped band  ignored. 
  FL means gapless Fermi liquid like phase.  }
      \label{fig:phase}
    \end{figure}
    Since we know that the instability will be resolved by developing a gap,   our recipe of the system for the new phase is 
    simply neglecting the k-gapped branch  of the spectral density. 
    Here we  speculate that we can classify the phases of the system with only other features in the spectral function.  
 The phase diagram  with  such scheme is drawn in Figure \ref{fig:phase} for $0.5<m<1.0$.    Dashed line implies that these `phases' are smoothly interpolated. The truth fate of the system should be calculated with explicit introduction of the non-homogeneity which will be much   heavier this this work.

\acknowledgments
 We thank   Matteo Baggioli for useful discussions. This  work is supported by Mid-career Researcher Program through the National Research Foundation of Korea grant No. NRF-2016R1A2B3007687.  YS is  supported  by Basic Science Research Program through NRF grant No. NRF-2016R1D1A1B03931443.
\bibliographystyle{JHEP}
 \bibliography{Refs_instability.bib}

\providecommand{\href}[2]{#2}\begingroup\raggedright\begin{thebibliography}{10}

\bibitem{WHANGBO_1991}
M.~H. Whangbo, E.~Canadell, P.~Foury and J.~P. Pouget, \emph{Hidden fermi
  surface nesting and charge density wave instability in low-dimensional
  metals}, \href{http://dx.doi.org/10.1126/science.252.5002.96}{\emph{Science}
  {\bf 252} (Apr, 1991) 96--98}.

\bibitem{Ling:2014saa}
Y.~Ling, C.~Niu, J.~Wu, Z.~Xian and H.-b. Zhang, \emph{{Metal-insulator
  Transition by Holographic Charge Density Waves}},
  \href{http://dx.doi.org/10.1103/PhysRevLett.113.091602}{\emph{Phys. Rev.
  Lett.} {\bf 113} (2014) 091602}, [\href{http://arxiv.org/abs/1404.0777}{{\tt
  1404.0777}}].

\bibitem{Amoretti:2017frz}
A.~Amoretti, D.~Aren, B.~Goutraux and D.~Musso, \emph{{Effective holographic
  theory of charge density waves}},
  \href{http://dx.doi.org/10.1103/PhysRevD.97.086017}{\emph{Phys. Rev.} {\bf
  D97} (2018) 086017}, [\href{http://arxiv.org/abs/1711.06610}{{\tt
  1711.06610}}].

\bibitem{Andrade:2017ghg}
T.~Andrade, A.~Krikun, K.~Schalm and J.~Zaanen, \emph{{Doping the holographic
  Mott insulator}},  \href{http://arxiv.org/abs/1710.05791}{{\tt 1710.05791}}.

\bibitem{Edalati:2010ge}
M.~Edalati, R.~G. Leigh, K.~W. Lo and P.~W. Phillips, \emph{{Dynamical Gap and
  Cuprate-like Physics from Holography}},
  \href{http://dx.doi.org/10.1103/PhysRevD.83.046012}{\emph{Phys. Rev.} {\bf
  D83} (2011) 046012}, [\href{http://arxiv.org/abs/1012.3751}{{\tt
  1012.3751}}].

\bibitem{Edalati:2010ww}
M.~Edalati, R.~G. Leigh and P.~W. Phillips, \emph{{Dynamically Generated Mott
  Gap from Holography}},
  \href{http://dx.doi.org/10.1103/PhysRevLett.106.091602}{\emph{Phys. Rev.
  Lett.} {\bf 106} (2011) 091602}, [\href{http://arxiv.org/abs/1010.3238}{{\tt
  1010.3238}}].

\bibitem{Liu:2009dm}
H.~Liu, J.~McGreevy and D.~Vegh, \emph{{Non-Fermi liquids from holography}},
  \href{http://dx.doi.org/10.1103/PhysRevD.83.065029}{\emph{Phys. Rev.} {\bf
  D83} (2011) 065029}, [\href{http://arxiv.org/abs/0903.2477}{{\tt
  0903.2477}}].

\bibitem{Gursoy:2011gz}
U.~Gursoy, E.~Plauschinn, H.~Stoof and S.~Vandoren, \emph{{Holography and ARPES
  Sum-Rules}}, \href{http://dx.doi.org/10.1007/JHEP05(2012)018}{\emph{JHEP}
  {\bf 05} (2012) 018}, [\href{http://arxiv.org/abs/1112.5074}{{\tt
  1112.5074}}].

\bibitem{Seo:2018hrc}
Y.~Seo, G.~Song, Y.-H. Qi and S.-J. Sin, \emph{{Mott transition with
  Holographic Spectral function}},  \href{http://arxiv.org/abs/1803.01864}{{\tt
  1803.01864}}.

\bibitem{sslee}
S.-S. Lee, \emph{{A Non-Fermi Liquid from a Charged Black Hole: A Critical
  Fermi Ball}}, \href{http://dx.doi.org/10.1103/PhysRevD.79.086006}{\emph{Phys.
  Rev.} {\bf D79} (2009) 086006}, [\href{http://arxiv.org/abs/0809.3402}{{\tt
  0809.3402}}].

\bibitem{Faulkner:2009wj}
T.~Faulkner, H.~Liu, J.~McGreevy and D.~Vegh, \emph{{Emergent quantum
  criticality, Fermi surfaces, and AdS(2)}},
  \href{http://dx.doi.org/10.1103/PhysRevD.83.125002}{\emph{Phys. Rev.} {\bf
  D83} (2011) 125002}, [\href{http://arxiv.org/abs/0907.2694}{{\tt
  0907.2694}}].

\bibitem{Faulkner:2011tm}
T.~Faulkner, N.~Iqbal, H.~Liu, J.~McGreevy and D.~Vegh, \emph{{Holographic
  non-Fermi liquid fixed points}},
  \href{http://dx.doi.org/10.1098/rsta.2010.0354}{\emph{Phil. Trans. Roy. Soc.}
  {\bf A 369} (2011) 1640}, [\href{http://arxiv.org/abs/1101.0597}{{\tt
  1101.0597}}].

\bibitem{Faulkner:2013bna}
T.~Faulkner, N.~Iqbal, H.~Liu, J.~McGreevy and D.~Vegh, \emph{{Charge transport
  by holographic Fermi surfaces}},
  \href{http://dx.doi.org/10.1103/PhysRevD.88.045016}{\emph{Phys. Rev.} {\bf
  D88} (2013) 045016}, [\href{http://arxiv.org/abs/1306.6396}{{\tt
  1306.6396}}].

\bibitem{Zaanen:2015oix}
J.~Zaanen, Y.-W. Sun, Y.~Liu and K.~Schalm, \emph{{Holographic Duality in
  Condensed Matter Physics}}.
\newblock Cambridge Univ. Press, 2015.

\bibitem{Hartnoll:2016apf}
S.~A. Hartnoll, A.~Lucas and S.~Sachdev, \emph{{Holographic quantum matter}},
  \href{http://arxiv.org/abs/1612.07324}{{\tt 1612.07324}}.

\bibitem{Cubrovic:2009ye}
M.~Cubrovic, J.~Zaanen and K.~Schalm, \emph{{String Theory, Quantum Phase
  Transitions and the Emergent Fermi-Liquid}},
  \href{http://dx.doi.org/10.1126/science.1174962}{\emph{Science} {\bf 325}
  (2009) 439--444}, [\href{http://arxiv.org/abs/0904.1993}{{\tt 0904.1993}}].

\bibitem{Cubrovic:2010bf}
M.~Cubrovic, J.~Zaanen and K.~Schalm, \emph{{Constructing the AdS Dual of a
  Fermi Liquid: AdS Black Holes with Dirac Hair}},
  \href{http://dx.doi.org/10.1007/JHEP10(2011)017}{\emph{JHEP} {\bf 10} (2011)
  017}, [\href{http://arxiv.org/abs/1012.5681}{{\tt 1012.5681}}].

\bibitem{Baggioli:2018vfc}
M.~Baggioli and K.~Trachenko, \emph{{Solidity of liquids: How Holography knows
  it}},  \href{http://arxiv.org/abs/1807.10530}{{\tt 1807.10530}}.

\bibitem{Baggioli:2018nnp}
M.~Baggioli and K.~Trachenko, \emph{{Maxwell interpolation and close
  similarities between liquids and holographic models}},
  \href{http://arxiv.org/abs/1808.05391}{{\tt 1808.05391}}.

\bibitem{blackburn2013x}
E.~Blackburn, J.~Chang, M.~H{\"u}cker, A.~Holmes, N.~B. Christensen, R.~Liang
  et~al., \emph{X-ray diffraction observations of a charge-density-wave order
  in superconducting ortho-ii yba 2 cu 3 o 6.54 single crystals in zero
  magnetic field}, {\emph{Physical review letters} {\bf 110} (2013) 137004}.

\bibitem{blanco2014resonant}
S.~Blanco-Canosa, A.~Frano, E.~Schierle, J.~Porras, T.~Loew, M.~Minola et~al.,
  \emph{Resonant x-ray scattering study of charge-density wave correlations in
  yba 2 cu 3 o 6+ x}, {\emph{Physical Review B} {\bf 90} (2014) 054513}.

\bibitem{Rice:1975aa}
T.~M. Rice, \emph{New mechanism for a charge-density-wave instability},
  \href{http://dx.doi.org/10.1103/PhysRevLett.35.120}{\emph{Physical Review
  Letters} {\bf 35} (1975) 120--123}.

\bibitem{Andrade:2011dg}
T.~Andrade and D.~Marolf, \emph{{AdS/CFT beyond the unitarity bound}},
  \href{http://dx.doi.org/10.1007/JHEP01(2012)049}{\emph{JHEP} {\bf 01} (2012)
  049}, [\href{http://arxiv.org/abs/1105.6337}{{\tt 1105.6337}}].

\bibitem{Amoretti_2018}
A.~Amoretti, D.~Are{\'a}n, B.~Gout{\'e}raux and D.~Musso, \emph{Effective
  holographic theory of charge density waves},
  \href{http://dx.doi.org/10.1103/physrevd.97.086017}{\emph{Physical Review D}
  {\bf 97} (Apr, 2018) }.

\bibitem{Amoretti_2018_2}
A.~Amoretti, D.~Are{\'a}n, B.~Gout{\'e}raux and D.~Musso, \emph{dc resistivity
  of quantum critical, charge density wave states from gauge-gravity duality},
  \href{http://dx.doi.org/10.1103/physrevlett.120.171603}{\emph{Physical Review
  Letters} {\bf 120} (Apr, 2018) }.

\bibitem{Musso:2018wbv}
D.~Musso, \emph{{Simplest phonons and pseudo-phonons in field theory}},
  \href{http://arxiv.org/abs/1810.01799}{{\tt 1810.01799}}.

\end{thebibliography}\endgroup

\end{document}